\documentclass[aps,prstab,groupedaddress,showpacs,showkeys,twocolumn]{revtex4-1}

\usepackage{graphicx} \usepackage{url} \usepackage{amsmath}
\usepackage{color}
\begin{document}

\title{A practical approach to extract symplectic transfer maps
  numerically for arbitrary magnetic elements}

\author{Yongjun Li} 

\affiliation{Brookhaven National Laboratory, Upton, NY-11973}

\author{Xiaobiao Huang}

\affiliation{SLAC National Accelerator Laboratory, Menlo Park,
  CA-94025}


\date{\today}

\begin{abstract}
  We introduce a practical approach to extract the symplectic transfer
  maps for arbitrary magnetic beam-line elements. Beam motion in particle
  accelerators depends on linear and nonlinear magnetic fields of the
  beam-line elements. These elements are usually modeled as magnetic
  multipoles with constant field strengths in the longitudinal direction
  (i.e., hard-edge model) in accelerator design and modeling codes. For
  magnets with complicated structures such as insertion devices or fields
  with significant longitudinal variation effects, the simplified models
  may not be sufficient to characterize beam dynamics behaviors
  accurately. A numerical approach has been developed to extract symplectic
  transfer maps from particle trajectory tracking simulation that uses
  magnetic field data provided by three-dimensional magnetic field
  modeling codes or experimental measurements.  The extracted transfer
  maps can be used in linear optics design and nonlinear dynamics
  optimization to achieve more realistic results.
\end{abstract}


\maketitle

\section{\label{intro}Introduction}
The designing and modeling of modern accelerators heavily rely on computer
codes such as MAD~\cite{mad}, Elegant~\cite{elegant}, Accelerator
Toolbox~\cite{AT}. In these codes, simplified magnet models are widely
used to calculate the transfer maps, which characterize the transportation
of charged particle through the magnets.  Typically magnets are described
as multipole fields about the design trajectory without longitudinal
variation.  Thus symplectic transfer maps can be easily obtained for such
so-called hard-edge models.  However, the field strengths of real magnets
always have longitudinal variations, at least in the transition regions at
both ends (i.e., fringe fields).  Although the magnetic field data in
three-dimensional (3D) space can usually be calculated with
electro-magnetic field solvers, it is rather difficult to extract the
transfer maps from the discrete field data directly.

Transfer maps characterizing the realistic fields are preferable in
magnetic element modeling. First, the fringe field effects can be
incorporated into the models. The consequences of the fringe fields are
particularly important for small rings with large acceptance and for beams
with large emittances as pointed out by Berz in Ref~\cite{Berz_fringe}.
It could also be critical for large rings with many magnets, such as the
proposed ultimate storage rings (USRs)~\cite{usr}, which typically
consists of many closely-packed strong magnets. In addition, the
cross-talk of the fringe fields of adjacent magnets could cause significant
differences between the real machines and the simplified
lattice models in such a scenario.  Second, special magnets, such as
insertion devices or combined-function dipoles with straight
geometry~\cite{spear_dipole}, can be accurately integrated into the 
lattice models.  By studying the discrepancies between the realistic
fields and the simplified magnet models, one can identify if each
individual model is sufficiently accurate from the view of linear and
nonlinear beam dynamics.

It is strongly desirable that transfer maps are symplectic because
symplecticity is essential in the study of long-term beam stability.  Lie
map, which ensures the symplecticity property of a Hamiltonian system, was
introduced by Dragt~\cite{Dragt_aip} into beam dynamics and is now widely
used in the accelerator community.  The purpose of this paper is to
present a general and practical approach to extract the Lie transfer map
of an arbitrary magnet by symplectifying the Taylor maps that are obtained
by fitting direct particle trajectory simulation data.  An alternative
method by surface field fitting to extract maps for straight-axis magnetic
elements can be found in Ref.~\cite{Mitchell}.

Our approach aims at constructing accurate lattice models by improving the
accuracy of individual magnet elements and potentially including the
cross-talk effects of adjacent magnets in the models.  Accurate lattice
models not only make nonlinear dynamics optimization results based on the
models more reliable, but also make machine commissioning easier since the
initial set-points of the magnets would be more accurate.

The paper is organized as follow: Section~\ref{map} reviews some basic
concepts of transfer maps for a Hamiltonian system and briefly introduces
the method established by Dragt and Finn~\cite{Dragt_Finn} to extract a
Lie map from a set of Taylor map series. In Section~\ref{procedure} we
discuss the procedure of map extraction step by step. Three detailed
examples are given in Section~\ref{application} as demonstrations of its
application. The paper concludes with a summary in
Section~\ref{conclusion}.

\section{\label{map}Transfer maps}
\subsection{Basic concepts}
We consider a dynamical system that is composed of a charged particle
moving through the field of a magnet. Ignoring radiation,
the transportation of the particle from the entrance face to the exit face
of the magnet can be represented by a transfer map in the six-dimensional
phase space of the canonical coordinates of the particle
\begin{equation}\label{coordinate}
X=[x,p_x,y,p_y,z,\delta]^T.
\end{equation}
Here superscript $T$ indicates vector transpose. As in most accelerator physics
literature~\cite{Lee_AP}, we use the path-length $s$ as the free variable. 
Therefore $(x, p_x)$, $(y,p_y)$ and
$(z=s-ct,\delta=\frac{P-P_0}{P_0})$ are three pairs of canonical
coordinates, which satisfy the Hamilton equations
\begin{equation}\label{H_eqs}
\dot{q_i}=\frac{\partial H}{\partial p_i},\: \dot{p_i}=-\frac{\partial
  H}{\partial q_i},
\end{equation}
where $q_i=x,y,z$ and $p_i=p_x,p_y,\delta$ for $i=1,2,3$,
respectively. The dots mean derivatives respective to the free variable
$s$ and $H$ is the Hamiltonian of the system.  The Lie transfer map for an
infinitesimal slice of the magnetic field can be expressed
as~\cite{Dragt_aip}
\begin{equation}\label{Lie}
\begin{aligned}
M(s\rightarrow{}s+\Delta{}s)&=e^{:-H(s)\Delta{}s:}=e^{:G:} \\
&=\left[1+:G:+\frac{:G:^2}{2!}+\cdots\right],
\end{aligned}
\end{equation}
where $G=-H(s)\Delta{}s$ is referred to as the Lie map generator and
$\Delta{}s$ is the length of the field slice.  Following
Dragt~\cite{Dragt_aip}, the Lie operator in Eq.~(\ref{Lie}) is defined to
signify the Poisson bracket, i.e.,
\begin{equation}
:f:g = [f,g]=\sum_{i=1}^3 \frac{\partial f}{\partial q_i}\frac{\partial
    g}{ \partial p_i} -\frac{\partial f}{\partial p_i}\frac{\partial
    g}{\partial q_i}
\end{equation}
for operator $:f:$ acting on
function $g$. The powers of an operator are defined as
\begin{equation}
\begin{aligned}
(:f:)^2g&=:f:(:f:g)=[f,[f,g]],\\
(:f:)^3g&=[f,[f,[f,g]]], \cdots, \: {\rm etc.}
\end{aligned}
\end{equation}

In general, the Lie map generators of beam-line elements can be expressed
as polynomials of the canonical coordinates
\begin{equation}
G=\sum_{abcdef}C_{abcdef}x^ap_x^by^cp_y^dz^e\delta^f. 
\end{equation}
For example, $G=-\frac{(p_x^2+p_y^2)L}{2}$ represents a drift space with
length $L$. 
It is convenient to express $G$ as~\cite{Chao_note}
\begin{equation}
G=\sum_{abcdef}C_{abcdef}|abcdef\rangle
\end{equation}
where $C_{abcdef}$ is the coefficient of the monomial term
\begin{equation}
|abcdef\rangle = x^ap_x^by^cp_y^dz^e\delta^f. 
\end{equation}

The transfer map of a beam-line composed of a series of magnet elements
can be obtained by joining the transfer maps of the individual magnets in
sequence~\cite{Chao_note}
\begin{equation}
M=e^{:f_1:}e^{:f_2:}\cdots{}e^{:f_n:},
\end{equation}
where $e^{:f_i:}$ is the transfer map of the $i$-th element. By means of
Lie algebra manipulations, such as similarity transformation and
Baker-Campbell-Hausdorff (BCH) theorem ~\cite{Dragt_aip,Chao_note}, one can
concatenate them into a single Lie map, which is known as the {\em
  one-turn-map} in accelerator physics literature if the beam-line is a 
closed loop.

Obviously, in order to obtain an accurate one-turn map to be used in
simulation and analysis, the transfer maps of the individual magnets need
to characterize the realistic fields precisely.

\subsection{Taylor map and Lie map}\label{DF}
The Taylor map approach plays an important role in the design of
charged-particle transport systems, such as linear accelerators,
synchrotron storage rings and spectrometers. The canonical coordinates of
a charged particle at the magnet exit are expressed as an expanded
multivariate Taylor power series of the coordinates at its
entrance~\cite{Brown}
\begin{equation}\label{Taylor}
X_{i,1}=\sum_{j=1}^{6}R_{ij}X_{j,0}+\sum_{j,k=1,j\le{}k}^{6}T_{ijk}X_{j,0}X_{k,0}+\cdots,
\end{equation}
where $R$ and $T$ are the $1^{\rm st}$- and $2^{\rm nd}$-order transfer map
coefficients, $X_{j,0}$ and $X_{j,1}$ are the $j$-th canonical coordinate
at the entrance and the exit, respectively.

Taylor maps can be very accurate in describing single pass trajectories.
But they are usually not symplectic. Therefore, they are not suitable for
the study of long-term stability in periodical structures, such as storage
rings. When simulating the trajectory of a relativistic particle under the
Lorentz force by solving the ordinary differential equations (ODE) , the
tracking result can be made very close to a symplectic transformation by
choosing small step sizes and tight convergence criteria with
non-symplectic integrators. Symplectic ODE integrators~\cite{Sanz_RK} are
also available. But they are usually implicit integrators and are
time-consuming.  Dragt and Finn have proved that~\cite{Dragt_Finn}, given
a symplectic Taylor map in the form of Eq.~(\ref{Taylor}), there exists an
infinite series of homogeneous multivariate polynomial Lie generators of
ascending orders, $G^{(2)}$, $G^{(3)}$, etc., such that the map
Eq.~(\ref{Taylor}) can be written in the form
\begin{equation}\label{DFmap}
X_{i,1}=\left[e^{:G^{(2)}:}e^{:G^{(3)}:}\cdots\right]X\bigg\vert_{X=X_{i,0}},
\end{equation}
where $G^{(i)},i\ge2$ is an $i$-th order homogeneous polynomial. This
method is known as the Dragt-Finn factorization. 

The factorization is realized by calculating the homogeneous generators
order by order. First the $2^{\rm nd}$-order generator can be derived from
the linear matrix (equivalent to the map $e^{:G^{(2)}:}$)
\begin{equation}\label{lin_mat}
R_{ij}=\frac{\partial{}X_{i,1}}{\partial{}X_{j,0}}. 
\end{equation}
To go to higher orders, the inverse map $e^{-:G^{(2)}:}$ is applied to the
Taylor map Eq.~(\ref{Taylor}) in order to obtain the residual map
$e^{-:G^{(2)}:} X_{1} - X_0 = X_r^{(2)} + O(3)$ (here $O(3)$ indicates
third or higher order terms) which should have no linear terms left
over. The quadratic terms in the residual map, $X_{r,i}^{(2)},
i=1,2,\cdots,6$, are the partial derivatives of an exact differential. The
$3^{\rm rd}$ order homogeneous generator $G^{(3)}$ is therefore given by a
path integral~\cite{Dragt_Finn}
\begin{eqnarray}\label{eq_G3}
 G^{(3)} = -\int^X \sum_{ij} X_{r,i}^{(2)} S_{ij} dX'_j,
\end{eqnarray} 
where $S$ is the $6\times6$ asymmetric symplecticity matrix with
$S_{ij}=[X_i, X_j]=:\!\!X_i\!\!:X_j$.  The same process can be repeated to
higher orders or when no higher order generator exists.  It can be
verified that the path integral for the $n^{\rm th}$ ($n\ge3$) generator is
given by
\begin{eqnarray}\label{eq_Gn}
G^{(n)} = -\frac1{n} \sum_{ij} X_{r,i}^{(n-1)} S_{ij} X_j,
\end{eqnarray} 
where $X_{r,i}^{(n-1)}$ is the $(n-1)^{\rm th}$ order polynomial terms in the
corresponding residual map $e^{-:G^{(n-1)}:}\cdots e^{-:G^{(2)}:} X_{1} -
X_0$.

If we can obtain near-symplectic Taylor maps by fitting simulated
multi-particle trajectories, it is straightforward to convert them to Lie
maps via Dragt-Finn factorization.  Taylor maps to arbitrary orders can
also be obtained with the technique of differential algebra
(DA)~\cite{Berz_DA}. To do so, the magnetic field must be given in
explicit functions of coordinates.  In most cases, it is difficult and
time-consuming to fit analytic functions to the discrete data on a 3D
grid.

\section{\label{procedure}Procedure of extracting transfer map}
For a given magnet, the procedure of extracting the Lie transfer map
through trajectory simulation is summarized in the following six steps:
\begin{enumerate}
 \item \emph{Obtaining the magnetic field data on a 3D grid.}  The size of
   the grid on which the three field components are given should be fine
   enough for 3D data interpolation during the trajectory simulation. If
   the grid is too sparse, the interpolated magnetic field might not be
   able to satisfy the Maxwell equations.

 \item \emph{Determining the reference orbit.} For straight line elements,
   such as quadrupoles, wigglers, etc., the reference orbit is a straight
   line passing through the magnet center. For curved magnets such as
   dipoles, the reference orbits may need to be numerically determined by
   finding a nominal particle trajectory that passes through the field
   under certain conditions. More details on curved reference orbit will
   be discussed in the second example in Section~\ref{application}. The
   coordinates of all particles at both the entrance and exit faces need
   to be converted relative to the local reference orbit.

 \item \emph{Implementing multi-particle trajectory simulation.}  The
   effects of the magnetic field to beam motion are sampled by tracking
   multiple particles through the field.  The initial phase space
   coordinates of the particles need to be populated evenly in the area of
   interest.  For example, if we are studying the dynamic aperture of a
   storage ring, the area in phase space should be larger than the
   expected dynamic aperture at the magnet location.  The number of
   coordinate values on each phase space direction should be sufficient in
   order to resolve the potential nonlinear effects to a certain order.
   In practice, multiple particle trajectory simulation is the most time
   consuming in this procedure (e.g., it takes ~20 hrs. for the 2-m magnet
   with a 1 mm step size in the second example in
   section~\ref{application} with serial computation.). However, this
   simulation only needs to be done once and the difficulty can be easily
   overcome by carrying out the simulation in parallel, if necessary.

 \item \emph{Fitting simulation results to Taylor maps.}  The Taylor map
   for particle coordinates from the entrance face to the exit face can be
   obtained by fitting the multi-particle tracking results with the least
   square minimization approach. The number of Taylor map coefficients can
   be calculated by
   \begin{equation}
   \sum_{k=1}^{\Omega}\frac{(n+k-1)!}{(n-1)k}\times{}n,
   \end{equation}
   where $n$ is the number of variables and $\Omega$ the order of the
   Taylor map. In order to ensure the minimization problem to be
   over-determined, the number of particles must be larger than the number
   of the coefficients. We can increase the particle number step by step
   until the Taylor map begins to converge within a tolerance. The order
   of the Taylor maps depends on the highest order of Lie generators user
   want to extract. For example, if one wants to generate up to the
   $4^{th}$ order Lie generators, the corresponding Taylor maps must be
   calculated to the $3^{rd}$ order, or even higher.
 
 \item \emph{Factorizing the Taylor map into a Lie map.}  The procedure of
   deriving a Lie map from Taylor maps is discussed in detail in
   Ref.~\cite{Dragt_Finn} and has been briefly recounted in
   Section~\ref{DF}.  It can be regarded as an automatic process of
   ``repairing'' the symplecticity property of the dynamical system. For
   each order of the factorization, after the inverse map is applied, the
   residual errors of this order should be exactly zeros because the
   corresponding generator is an exact path integral for a symplectic map.
   By dropping off the residual errors, the slight error of symplecticity
   due to, e.g. the numerical errors in 3D data interpolation, the
   implementation of non-symplectic integrators in solving the ODEs, etc.,
   are rounded off.  On the other hand, the residual errors serve as an
   indication of the symplectic quality of the Taylor map obtained from
   fitting the trajectory simulation data. The factorization process can
   stop at a certain order, or when no higher order generators exist.
 
 \item \emph{Validating the Lie map by comparing to the simulation data.}
   The accuracy of the Lie map can be checked by evaluating the
   transportation of individual particles using Eqs.~(\ref{Lie}) and
   (\ref{DFmap}), and comparing the results with trajectory tracking
   results. The comparison indicates the quality of the Lie map, and may
   also suggests whether one needs to push the factorization to higher
   orders in order to get a precise transfer map.
\end{enumerate}

The Lie map obtained with the above procedure serves as an accurate and
concise description of the element.  It can be used for non-symplectic
particle tracking with Eqs.~(\ref{DFmap}) and (\ref{Lie}) while truncating
at the desired order.  It is also possible to implement symplectic
tracking with the map. One symplectic tracking method is achieved by splitting
the exponential map of polynomials into a series of
exponential maps of monomials with the BCH
formulas~\cite{Chao_note}.  Each monomial exponential map can be
evaluated symplecticly.

In the next section, three applications are given for the purpose of
demonstration of the method.
 
\section{\label{application}Applications}
\subsection{\label{quad}Fringe field of a quadrupole }
Studies of magnet fringe field effects can be found in many accelerator
literature
~\cite{Enge_fringe,Berz_fringe,Forest_fringe,Zimmermann_fringe}. The
purpose of this example is to benchmark our approach against the
well-established COSY-Infinity code~\cite{Berz_cosy}, which provides an
accurate soft fringe model for quadrupoles based on the differential
algebra (DA) technique~\cite{Berz_DA}.

\begin{figure}[!ht]
  \centering
  \includegraphics[width=\columnwidth]{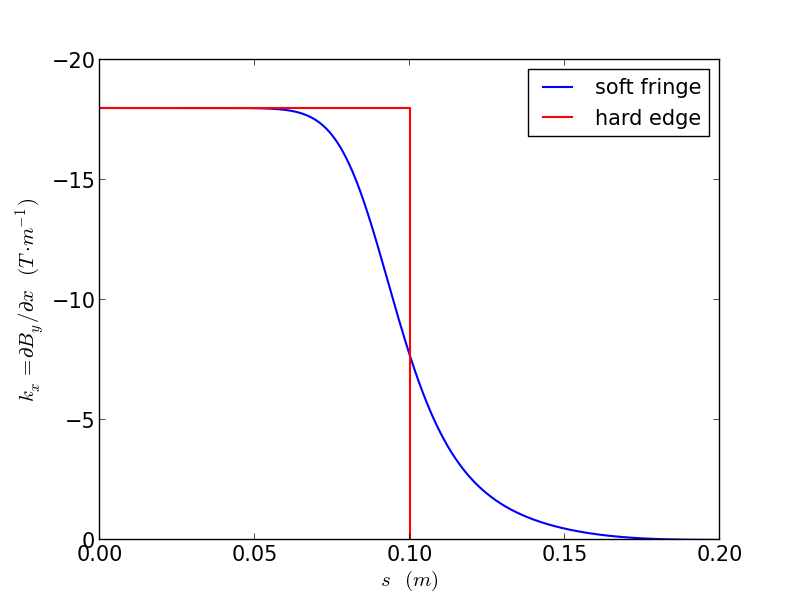}
  \caption{\label{soft_quad_profile}A soft fringe field profile described
    by Enge function}
\end{figure}

Consider a quadrupole with a soft fringe field at its exit as illustrated
in FIG.~\ref{soft_quad_profile}. The nominal field gradient is
$\frac{\partial{}B_y}{\partial{}x}=-18 $~T/m and the effective
length $L_{\rm eff}=0.1$~m. The fringe field fall-off profile is described
by the Enge function with six parameters~\cite{Berz_cosy}
\begin{equation}\label{Enge}
F(s)=\frac{1}{1+\mbox{exp}(\sum_{i=1}^6a_i\frac{(s-s_0)^{i-1}}{D})},
\end{equation}
where $D=0.05$~m is the full aperture and $s_0=0.1$~m is the effective field
boundary. We calculate the quadrupole field on a 3D grid based on
Eq.~(\ref{Enge}) with the default Enge coefficients in COSY, $a_i$= 0.296471,
  4.533219, $-2.270982$, 1.068627, $-0.036391$, 0.022261 for $i=1$ to 6.

\begin{figure}[!ht]
  \centering
  \includegraphics[width=\columnwidth]{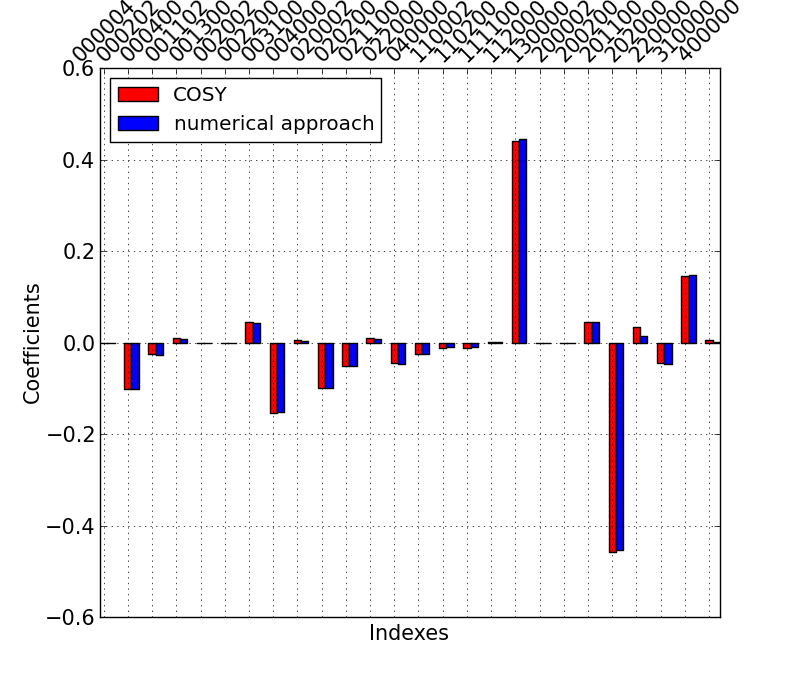}
  \caption{\label{soft_quad_cmp_cosy}Comparison of the nonvanishing
    $4^{\rm th}$-order homogeneous polynomial coefficients between COSY
    and the numerical approach.}
\end{figure}

Here we want to obtain the transfer map for a 3 GeV electron beam by
following the procedure summarized in section~\ref{procedure}. The
reference orbit is the straight line passing through the quadrupole
center. Multiple particles with different initial coordinates are tracked
through the magnet by using the Runge-Kutta ODE integration method to
solve the equations of motion under the Lorentz force. The longitudinal
free coordinate $s$ must cover the fringe field region for the trajectory
simulation. All coordinates at both the entrance and the exit are
recorded, from which the Taylor transfer map is extracted. Finally the Lie
transfer map is factorized from the Taylor map. The comparison of the
nonvanishing $4^{\rm th}$ order homogeneous polynomial coefficients
between the two approaches (FIG.~\ref{soft_quad_cmp_cosy}) shows that they
agree with each other very well.

The obtained transfer map has been validated by evaluating
Eqs.~(\ref{DFmap}) and (\ref{Lie}) for each simulated particle. By using
the first three generators $G^{(2-4)}$ and truncating the exponential map
Eq.~(\ref{Lie}) at the $5^{\rm th}$-order, the standard deviations (RMS) of
the discrepancy between the direct trajectory simulation and the map
transportation are found to be less than $1\times10^{-7}$~m or rad for the
phase space coordinates.

\subsection{\label{bend}Combined-function dipole with a Cartesian gradient}
\begin{figure}[!ht]
  \centering
  \includegraphics[width=\columnwidth]{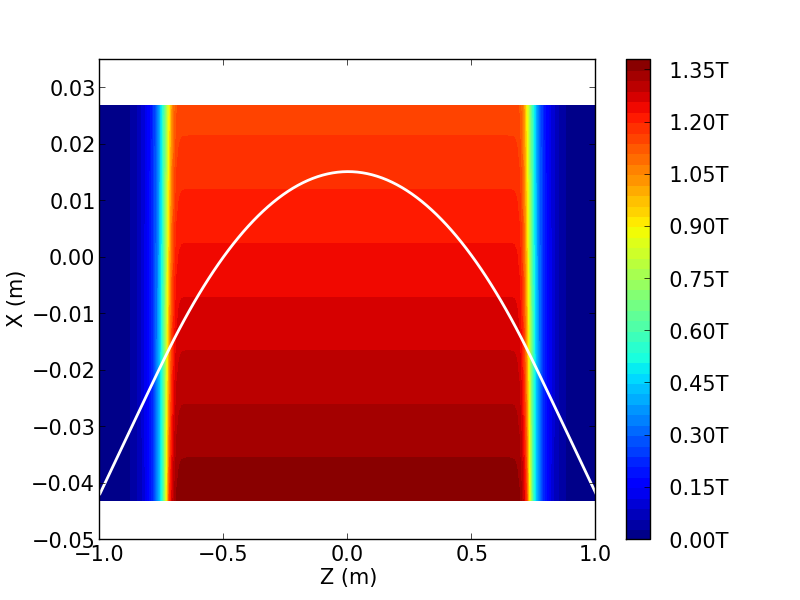}
  \caption{\label{spear_dipole_field_traj}Contour of the vertical field
    component of a standard SPEAR3 dipole in the mid-plane. The white line
    is the reference orbit obtained by simulation.}
\end{figure}

The dipole magnets of SPEAR3 and a few other storage rings are
combined-function magnets built with a straight
geometry~\cite{spear_dipole,speardipole_Yoon}. Therefore, the magnetic
field seen by a particle and its curvature of trajectory vary along the path
inside the magnet (see FIG.~\ref{spear_dipole_field_traj}), in addition to
variations in the fringe regions.  
A direct trajectory tracking has been used to study
the linear and nonlinear effects of this type of
magnets~\cite{Huang_IPAC10}. But the method is non-symplectic and 
computationally very slow and is hence not suitable for beam dynamics optimization.

The magnetic field for the SPEAR3 dipole in this study is based on 
the analytic field model in Ref.~\cite{Huang_IPAC10} which was derived from 
magnetic field measurements on the mid-plane.   

First, the reference orbit is obtained by launching a particle on the
mid-plane with the nominal energy (3 GeV) and the nominal entrance angle
(half of the nominal deflection angle) and varying its launching position
until a symmetric trajectory is found.  The particle trajectories need to
start and end in field-free locations.  The reference trajectory is shown
in FIG.~\ref{spear_dipole_field_traj}.  Multiple particle trajectories
about the reference trajectory are simulated with tracking.  The dynamic
aperture at the location of the dipole is less than $1$~cm for both
transverse planes because of the small horizontal beta function
($\beta_x<2$~m) at the location and small vertical physical apertures at
the insertion devices.  Therefore the input position coordinates for
trajectory simulation are populated evenly within a $1\:{\rm cm}\times
1\:{\rm cm}$ box in the $x$-$y$ plane.

\begin{figure}[!ht]
  \centering
  \includegraphics[width=\columnwidth]{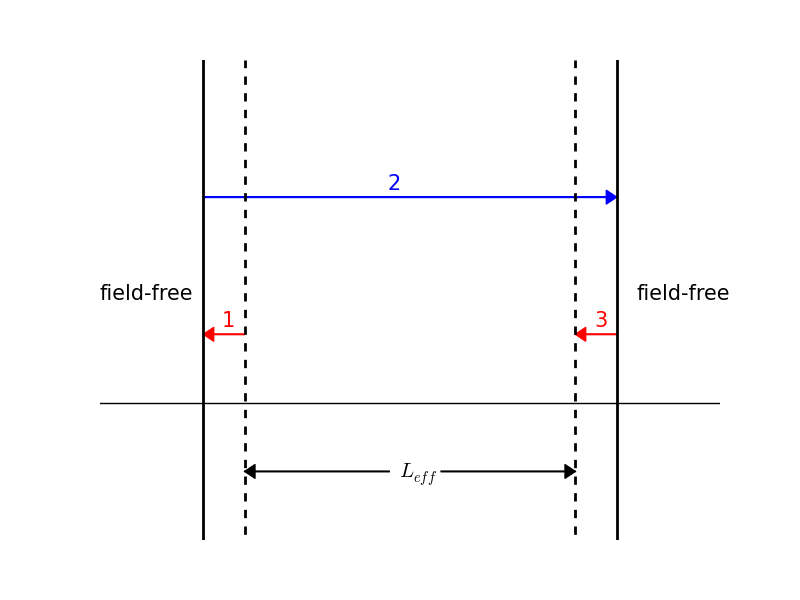}
  \caption{\label{dipole_effective_length}Illustration of the procedure to
    obtain the transfer map for an equivalent hard-edge model by adding
    two virtual negative drifts (the step 1 and 3 shown as red arrows) at
    the ends.}
\end{figure}

In accelerator lattice design, it is preferable to model a magnet
compactly with hard edges at an appropriate effective length.  This can be
realized by adding two virtual negative length drifts on both sides, as
illustrated in FIG.~\ref{dipole_effective_length}.  Starting from the
entrance of the hard-edge model, the particles first drift backward to the
field-free region (step 1). Then the particles are tracked through the
magnetic field region (step 2).  Finally at the exit side the particles
again drift backward from the field-free edge to the hard-edge boundary
(step 3).  Because numerical tracking is done on the Cartesian
coordinates, coordinate transformations are needed between the usual
Frenet-Serret coordinate system and the Cartesian coordinate system at the
hard edge boundaries of both ends.

With the tracking results from the above setup and the procedure described
in section~\ref{procedure}, we calculated the transfer map for an
equivalent hard-edge model for the dipole that includes the fringe field
effects.  This map can be incorporated into a lattice model for beam
dynamics analysis. The linear transfer matrix for the SPEAR3 dipole is
listed in Table~\ref{spear3_dipole_matrix}, and the first few monomial
coefficients in generators $G^{(3-4)}$ are listed in descending order in
Table~\ref{spear3_dipole_g34}. The RMS values of discrepancy between the
trajectory simulation and the map transportation are around
$2.5\times10^{-5}$~m or rad if the three generators $G^{(2-4)}$ are used
for evaluation. The discrepancy will be further reduced to around
$1.1\times10^{-5}$~m or rad if the generator $G^{(5)}$ is implemented.

A usual hard-edge sector dipole model can be derived from the magnetic
field profile of the SPEAR3 dipole. The effective length is 1.50694 m, the
bending angle is $\pi/17$ and the focusing strength is
$K_1=-0.3117031$~${\rm{}m}^{-2}$.  The difference between this simplified
model and the Lie map obtained numerically can be accounted for with a
virtual thin-lens corrector element attached at the end of the sector
dipole model.  The linear transfer matrix for this corrector is shown in
Table~\ref{spear3_dipole_matrix_diff}, which represents linear errors of
the best sector dipole model.  Incorporating corrector elements like this
into the lattice model would improve model accuracy.

\begin{table}[ht]
\caption{6D Linear transfer matrix for a standard SPEAR3 dipole (in SI
  units, same below)}
\label{spear3_dipole_matrix}
\begin{tabular}{rrrrrr}
\hline\hline
\tt{1.37520}&\tt{1.68361}&\tt{0}&\tt{0}&\tt{0}&\tt{0.14747}\\
\tt{0.52933}&\tt{1.37521}&\tt{0}&\tt{0}&\tt{0}&\tt{0.20804}\\
\tt{0}&\tt{0}&\tt{0.65202}&\tt{1.33404}&\tt{0}&\tt{0}\\
\tt{0}&\tt{0}&\tt{-0.43092}&\tt{0.65202}&\tt{0}&\tt{0}\\
\tt{-0.20804}&\tt{-0.14747}&\tt{0}&\tt{0}&\tt{1}&\tt{-0.00905}\\
\tt{0}&\tt{0}&\tt{0}&\tt{0}&\tt{0}&\tt{1}\\ \hline
\end{tabular}
\end{table}

\begin{table}[ht]
\caption{A correction transfer matrix at the exit end of the sector dipole
  model to account for the difference between such a model and the Lie map
  for a standard SPEAR3 dipole}
\label{spear3_dipole_matrix_diff}
\begin{tabular}{rrrrrr}
\hline\hline
\tt{0.99898}&\tt{0.00246}&\tt{0}&\tt{0}&\tt{0}&\tt{-0.00013}\\
\tt{-0.00089}&\tt{1.00102}&\tt{0}&\tt{0}&\tt{0}&\tt{0.00003}\\
\tt{0}&\tt{0}&\tt{0.99947}&\tt{-0.00092}&\tt{0}&\tt{0}\\
\tt{0}&\tt{0}&\tt{-0.00023}&\tt{1.00053}&\tt{0}&\tt{0}\\
\tt{-0.00003}&\tt{0.00013}&\tt{0}&\tt{0}&\tt{1}&\tt{-0.00015}\\
\tt{0}&\tt{0}&\tt{0}&\tt{0}&\tt{0}&\tt{1}\\ \hline
\end{tabular}
\end{table}

\begin{center}
\begin{table}[htb]
\caption{Part of the $3^{\rm rd}$- and $4^{\rm th}$-order generator coefficients
  for the SPEAR3 dipole}
\label{spear3_dipole_g34}
\begin{tabular}{p{.5cm}p{1cm}p{.5cm}p{1cm}p{.5cm}p{1cm}r}
\hline\hline
a&b&c&d&e&f&Coefficient\\
\hline
0&2&0&0&0&1&      1.13814\\
0&0&0&2&0&1&      0.75041\\
1&1&0&0&0&1&    - 0.49203\\
1&2&0&0&0&0&     -0.38990\\
0&1&1&1&0&0&      0.37658\\
$\cdots$&&&&&&\\
\hline
0&2&0&0&0&2&     -1.45884\\
1&3&0&0&0&0&      1.13290\\
2&2&0&0&0&0&     -0.91574\\
0&0&0&2&0&2&     -0.74524\\
0&4&0&0&0&0&     -0.67212\\
$\cdots$&&&&&&\\
\hline\hline
\end{tabular}
\end{table}
\end{center}

\subsection{\label{id}Insertion device integration}
Insertion devices (ID), such as wigglers and undulators, are the main
x-ray sources in modern storage ring light sources. Usually their first
and second field integrals are required to be zeros so that they can be
regarded as straight beam-line elements, although the real reference
orbits inside the magnet bodies actually wiggle, or spiral around a
straight line.

Several symplectic integrators~\cite{Elleaume_kickmap,Wu_id,Bahrdt_GF} are
available for particle tracking through IDs. But in some cases, it is
desirable to have explicit transfer maps. For example, by integrating IDs
into the one-turn map of the storage ring, their contributions to the
nonlinear resonance driving terms (NRDT) can be calculated quantitatively.

We use an elliptical polarized undulator (EPU) of the NSLS-II NEXT
project~\cite{nsls2_NEXT} as an example to illustrate the application of
our method to IDs.  The magnetic field for the vertical polarized mode of
this device is calculated using the code RADIA~\cite{Chubar_radia}. The
first few main generator coefficients of $G^{(4)}$ for one period of the
device are listed in descending order in Table ~\ref{nsls2_epu49_g4}. As
expected, the EPU has some very weak sextupole-like components in
$G^{(3)}$ (not listed in the table). But the dominant contributions are
from the vertical octupole-like components in $G^{(4)}$. These terms
contribute directly to the $2^{\rm nd}$ order driving terms, and eventually
affect the storage ring dynamic aperture. Because the coefficients given
in Table~\ref{nsls2_epu49_g4} are for only one period and an EPU typically
have tens of periods, the device could have a significant impact on the
nonlinear dynamics of the storage ring. Therefore it is necessary to
incorporate IDs into the storage ring lattice optimization.

\begin{center}
\begin{table}[htb]
\caption{Part of the $4^{\rm th}$-order generator coefficients
  for one of the NSLS-II EPUs}
\label{nsls2_epu49_g4}
\begin{tabular}{p{.5cm}p{1cm}p{.5cm}p{1cm}p{.5cm}p{1cm}r}
\hline\hline
a&b&c&d&e&f&Coefficient\\
\hline
0 & 0 & 4 & 0 & 0 & 0 &      -1.60204\\
0 & 0 & 1 & 3 & 0 & 0 &       1.07844\\
1 & 1 & 2 & 0 & 0 & 0 &       0.10163\\
2 & 0 & 2 & 0 & 0 & 0 &       0.06378\\
2 & 0 & 1 & 1 & 0 & 0 &       0.05501\\
1 & 3 & 0 & 0 & 0 & 0 &      -0.04480\\
$\cdots$&&&&&&\\
\hline\hline
\end{tabular}
\end{table}
\end{center}

\begin{figure}[!ht]
  \centering
  \includegraphics[width=\columnwidth]{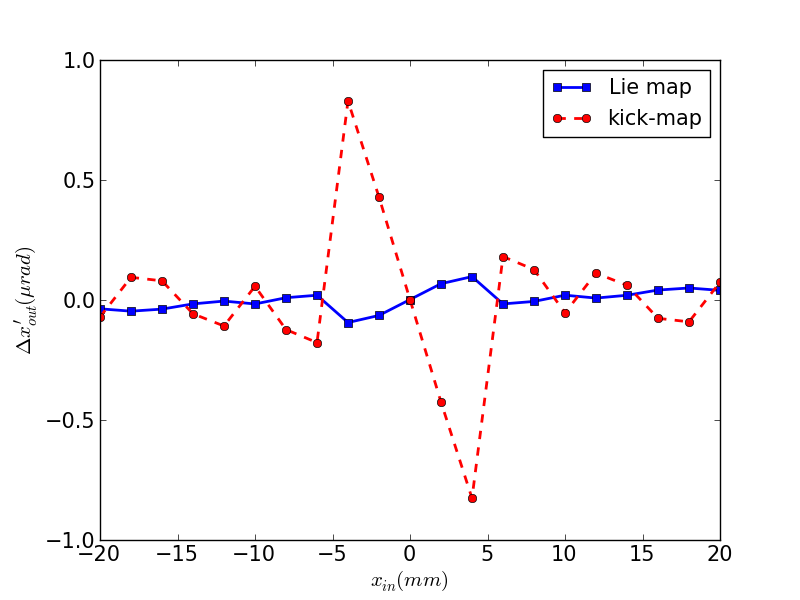}
  \caption{\label{id_rk_km_tm} Discrepancy of the horizontal exit angles
    between the Lie generators (transfer map) and the kick-map generator
    after passing through one period of the EPU at different initial
    coordinates. The Runge-Kutta integrator's result is used as the
    reference. The vertical offset is fixed at 2~mm.}
\end{figure}

It is interesting to compare the Lie transfer map we obtained against the
direct trajectory simulation and the kick-map~\cite{Elleaume_kickmap}
calculated by RADIA~\cite{Chubar_radia}. We chose a set of particles with
different initial horizontal offsets within the $[-2,2]$~cm range and a
fixed vertical offset at $y=2$~mm and transported them through one period
of the EPU with the Runge-Kutta integrator, the Lie transfer map and the
kick-map integrator, respectively.  Using the results of the Runge-Kutta
integrator as the reference, the discrepancies of the other two
integrators with respect to the reference at different initial horizontal
coordinates are shown in FIG.~\ref{id_rk_km_tm}. Although both the
kick-map integrator and Lie generator can achieve results with an error of
$\Delta{}x^{'}_{max}<{}1.0\mu{}rad$, the Lie transfer map has better
performance in terms of absolute errors and the smoothness of the error
curve.  It is not straight-forward to obtain the Lie map from the 3D field
data. But it is worth doing, not only because the Lie map can generate
better results in tracking, but also because it can be directly used in
calculating the contribution to the resonance driving terms.

To calculate the contributions to the $2^{\rm nd}$-order driving terms from
the extracted Lie transfer map, we need to expand the Lie generator
$G^{(4)}$ to a polynomial in the betatron oscillation resonance
basis~\cite{Chao_note}
\begin{equation}
h_{x,y}^\pm=\sqrt{2J_{x,y}}e^{\pm{}i\phi_{x,y}}=(\bar{x},\bar{y})\mp{}i\bar{p}_{x,y},
\end{equation}
where $J_{x,y}$ and $\phi_{x,y}$ are action-angle variables,
$\bar{x},\bar{y}$ and $\bar{p}_{x,y}$  normalized coordinates which are related with the
Courant-Snyder parameters $\alpha_{x,y},\beta_{x,y}$ at the location of the ID through 
\begin{equation}\label{normalized_x_p}
\left[
\begin{array}{c}
\bar{x}\\
\bar{p}_x
\end{array}
\right]=
\left[
\begin{array}{cc}
\frac{1}{\sqrt{\beta_x}} & 0\\
\frac{\alpha_x}{\sqrt{\beta_x}} & \sqrt{\beta}_x\\
\end{array}
\right]
\left[
\begin{array}{c}
x\\
p_x
\end{array}
\right],
\end{equation}
and likewise for the $y$-plane. 
The new generator polynomial in the resonance basis can be expressed as
\begin{equation}\label{g4}
G^{(4)}=\sum_{\substack{i,j,k,l,m=0\\ i+j+k+l+m=4}}C_{ijkl,m}|ijkl,m\rangle,
\end{equation}
where the monomial signifies  
\begin{equation}\label{ijklm}
|ijkl,m\rangle=(h_x^+)^i(h_x^-)^j(h_y^+)^k(h_y^-)^l\delta^m{}.
\end{equation}
Here only five power indexes appear in Eqs.~(\ref{g4}) and (\ref{ijklm})
because the $z=s-ct$ coordinate has no dynamic effect in an ID. It is well-known that $C_{1111,0}$,
$C_{2200,0}$ and $C_{0022,0}$ are related to the three first order
tune-shift-with-amplitude coefficients, which are important in dynamic
aperture optimization.

\section{\label{conclusion}Conclusion}
We have introduced a general method to extract a symplectic transfer map
for an arbitrary beam-line element from direct trajectory simulation
through its magnetic field. A Taylor map is first obtained by fitting the
multi-particle trajectory tracking data, which is then turned into a Lie map
with the Dragt-Finn factorization~\cite{Dragt_Finn}.  The method is
demonstrated with applications to three practical examples, including the
fringe field effect of quadrupoles, dynamic effects of a straight-geometry
combined-function dipole and an insertion device.

The Lie map obtained with our method can be used in lattice models to
study linear and nonlinear beam dynamics effects of various beam-line
elements that may have been left out in existing codes, such as
combined-function dipoles on straight-geometry that are used on a few
storage ring light sources.  Another potential application is the study of 
the cross-talk effect between adjacent beam-line elements. This could be
especially important for future ultimate storage rings. The 
proposed USR lattice designs are similar to that of the
MAX-IV~\cite{Leemann_max4} storage ring in which many small magnets are
closely packed. By applying our numerical approach to a sequence of
magnets to extract a combined symplectic transfer map, we may
directly study the beam dynamics with the cross-talk effects included, or
separate the cross-talking effects from each individual magnet by
introducing correction maps in between.

\begin{acknowledgments}
  We thank Prof. Dragt and his student Dr. Mitchell for many useful
  discussions, Prof. Berz and his student Dr. H. Zhang for help in using
  COSY-Infinity, and Dr. Kitegi for providing the NSLS-II EPU field data.
  We are also grateful to J. Safranek and S. Krinsky for their pertinent
  comments on an earlier version of the manuscrip. Work at Brookhaven
  National Laboratory was supported by the U.S. Department of Energy,
  Office of Science, Office of Basic Energy Sciences, under Contract
  No. DE-AC02-98CH10886.  Work at SLAC National Accelerator Laboratory was
  supported by the U.S. Department of Energy, Office of Science, Office of
  Basic Energy Sciences, under Contract No.  DE-AC02-76SF00515.
\end{acknowledgments}

\bibliography{map_ref}

\begin{thebibliography}{26}%
\makeatletter
\providecommand \@ifxundefined [1]{%
 \@ifx{#1\undefined}
}%
\providecommand \@ifnum [1]{%
 \ifnum #1\expandafter \@firstoftwo
 \else \expandafter \@secondoftwo
 \fi
}%
\providecommand \@ifx [1]{%
 \ifx #1\expandafter \@firstoftwo
 \else \expandafter \@secondoftwo
 \fi
}%
\providecommand \natexlab [1]{#1}%
\providecommand \enquote  [1]{``#1''}%
\providecommand \bibnamefont  [1]{#1}%
\providecommand \bibfnamefont [1]{#1}%
\providecommand \citenamefont [1]{#1}%
\providecommand \href@noop [0]{\@secondoftwo}%
\providecommand \href [0]{\begingroup \@sanitize@url \@href}%
\providecommand \@href[1]{\@@startlink{#1}\@@href}%
\providecommand \@@href[1]{\endgroup#1\@@endlink}%
\providecommand \@sanitize@url [0]{\catcode `\\12\catcode `\$12\catcode
  `\&12\catcode `\#12\catcode `\^12\catcode `\_12\catcode `\%12\relax}%
\providecommand \@@startlink[1]{}%
\providecommand \@@endlink[0]{}%
\providecommand \url  [0]{\begingroup\@sanitize@url \@url }%
\providecommand \@url [1]{\endgroup\@href {#1}{\urlprefix }}%
\providecommand \urlprefix  [0]{URL }%
\providecommand \Eprint [0]{\href }%
\providecommand \doibase [0]{http://dx.doi.org/}%
\providecommand \selectlanguage [0]{\@gobble}%
\providecommand \bibinfo  [0]{\@secondoftwo}%
\providecommand \bibfield  [0]{\@secondoftwo}%
\providecommand \translation [1]{[#1]}%
\providecommand \BibitemOpen [0]{}%
\providecommand \bibitemStop [0]{}%
\providecommand \bibitemNoStop [0]{.\EOS\space}%
\providecommand \EOS [0]{\spacefactor3000\relax}%
\providecommand \BibitemShut  [1]{\csname bibitem#1\endcsname}%
\let\auto@bib@innerbib\@empty
\bibitem [{\citenamefont {Grote}\ and\ \citenamefont {Schmidt}(2003)}]{mad}%
  \BibitemOpen
  \bibfield  {author} {\bibinfo {author} {\bibfnamefont {H.}~\bibnamefont
  {Grote}}\ and\ \bibinfo {author} {\bibfnamefont {F.}~\bibnamefont
  {Schmidt}},\ }\href@noop {} {\  (\bibinfo {year} {2003})},\ \bibinfo {note}
  {{CERN-AB-2003-024-ABP}}\BibitemShut {NoStop}%
\bibitem [{\citenamefont {Borland}(2000)}]{elegant}%
  \BibitemOpen
  \bibfield  {author} {\bibinfo {author} {\bibfnamefont {M.}~\bibnamefont
  {Borland}},\ }\href@noop {} {\  (\bibinfo {year} {2000})},\ \bibinfo {note}
  {{Advanced Photon Source LS-287}}\BibitemShut {NoStop}%
\bibitem [{\citenamefont {Terebilo}(2001)}]{AT}%
  \BibitemOpen
  \bibfield  {author} {\bibinfo {author} {\bibfnamefont {A.}~\bibnamefont
  {Terebilo}},\ }\href@noop {} {\  (\bibinfo {year} {2001})},\ \bibinfo {note}
  {{SLAC-PUB-8732}}\BibitemShut {NoStop}%
\bibitem [{\citenamefont {Berz}\ \emph {et~al.}(2000)\citenamefont {Berz},
  \citenamefont {Erdelyi},\ and\ \citenamefont {Makino}}]{Berz_fringe}%
  \BibitemOpen
  \bibfield  {author} {\bibinfo {author} {\bibfnamefont {M.}~\bibnamefont
  {Berz}}, \bibinfo {author} {\bibfnamefont {B.}~\bibnamefont {Erdelyi}}, \
  and\ \bibinfo {author} {\bibfnamefont {K.}~\bibnamefont {Makino}},\
  }\href@noop {} {\bibfield  {journal} {\bibinfo  {journal}
  {Nucl.Instrum.Meth.}\ }\textbf {\bibinfo {volume} {A472}},\ \bibinfo {pages}
  {533} (\bibinfo {year} {2000})}\BibitemShut {NoStop}%
\bibitem [{\citenamefont {Reich}(2013)}]{usr}%
  \BibitemOpen
  \bibfield  {author} {\bibinfo {author} {\bibfnamefont {E.~S.}\ \bibnamefont
  {Reich}},\ }\href@noop {} {\bibfield  {journal} {\bibinfo  {journal}
  {Nature}\ }\textbf {\bibinfo {volume} {501}},\ \bibinfo {pages} {148}
  (\bibinfo {year} {2013})}\BibitemShut {NoStop}%
\bibitem [{\citenamefont {Corbett}\ \emph {et~al.}(1999)\citenamefont
  {Corbett}, \citenamefont {Dell'Orco}, \citenamefont {Nosochkov},\ and\
  \citenamefont {Tanabe}}]{spear_dipole}%
  \BibitemOpen
  \bibfield  {author} {\bibinfo {author} {\bibfnamefont {J.}~\bibnamefont
  {Corbett}}, \bibinfo {author} {\bibfnamefont {D.}~\bibnamefont {Dell'Orco}},
  \bibinfo {author} {\bibfnamefont {Y.}~\bibnamefont {Nosochkov}}, \ and\
  \bibinfo {author} {\bibfnamefont {J.}~\bibnamefont {Tanabe}},\ }\href@noop {}
  {\  (\bibinfo {year} {1999})},\ \bibinfo {note} {{SLAC-PUB-8234}}\BibitemShut
  {NoStop}%
\bibitem [{\citenamefont {Dragt}(1982)}]{Dragt_aip}%
  \BibitemOpen
  \bibfield  {author} {\bibinfo {author} {\bibfnamefont {A.}~\bibnamefont
  {Dragt}},\ }\href@noop {} {\bibfield  {journal} {\bibinfo  {journal} {AIP
  Conf.Proc.}\ }\textbf {\bibinfo {volume} {87}},\ \bibinfo {pages} {147}
  (\bibinfo {year} {1982})}\BibitemShut {NoStop}%
\bibitem [{\citenamefont {Mitchell}\ and\ \citenamefont
  {Dragt}(2010)}]{Mitchell}%
  \BibitemOpen
  \bibfield  {author} {\bibinfo {author} {\bibfnamefont {C.}~\bibnamefont
  {Mitchell}}\ and\ \bibinfo {author} {\bibfnamefont {A.}~\bibnamefont
  {Dragt}},\ }\href@noop {} {\bibfield  {journal} {\bibinfo  {journal}
  {Phys.Rev.ST Accel.Beams}\ }\textbf {\bibinfo {volume} {13}},\ \bibinfo
  {pages} {064001} (\bibinfo {year} {2010})}\BibitemShut {NoStop}%
\bibitem [{\citenamefont {Dragt}\ and\ \citenamefont
  {Finn}(1976)}]{Dragt_Finn}%
  \BibitemOpen
  \bibfield  {author} {\bibinfo {author} {\bibfnamefont {A.}~\bibnamefont
  {Dragt}}\ and\ \bibinfo {author} {\bibfnamefont {J.}~\bibnamefont {Finn}},\
  }\href@noop {} {\bibfield  {journal} {\bibinfo  {journal} {J.Math.Phys.}\
  }\textbf {\bibinfo {volume} {17}},\ \bibinfo {pages} {2215} (\bibinfo {year}
  {1976})}\BibitemShut {NoStop}%
\bibitem [{\citenamefont {Lee}(1999)}]{Lee_AP}%
  \BibitemOpen
  \bibfield  {author} {\bibinfo {author} {\bibfnamefont {S.~Y.}\ \bibnamefont
  {Lee}},\ }\href@noop {} {\emph {\bibinfo {title} {{Accelerator physics}}}}\
  (\bibinfo  {publisher} {World Scientific},\ \bibinfo {year}
  {1999})\BibitemShut {NoStop}%
\bibitem [{\citenamefont {Chao}(2002)}]{Chao_note}%
  \BibitemOpen
  \bibfield  {author} {\bibinfo {author} {\bibfnamefont {A.}~\bibnamefont
  {Chao}},\ }\href@noop {} {\  (\bibinfo {year} {2002})},\ \bibinfo {note}
  {{SLAC-PUB-9574}}\BibitemShut {NoStop}%
\bibitem [{\citenamefont {Brown}(1968)}]{Brown}%
  \BibitemOpen
  \bibfield  {author} {\bibinfo {author} {\bibfnamefont {K.}~\bibnamefont
  {Brown}},\ }\href@noop {} {\  (\bibinfo {year} {1968})},\ \bibinfo {note}
  {{SLAC-R-075, SLAC-75}}\BibitemShut {NoStop}%
\bibitem [{\citenamefont {Sanz-Serna}(1988)}]{Sanz_RK}%
  \BibitemOpen
  \bibfield  {author} {\bibinfo {author} {\bibfnamefont {J.}~\bibnamefont
  {Sanz-Serna}},\ }\href@noop {} {\bibfield  {journal} {\bibinfo  {journal}
  {{BIT}}\ }\textbf {\bibinfo {volume} {28}},\ \bibinfo {pages} {877} (\bibinfo
  {year} {1988})}\BibitemShut {NoStop}%
\bibitem [{\citenamefont {Berz}(1989)}]{Berz_DA}%
  \BibitemOpen
  \bibfield  {author} {\bibinfo {author} {\bibfnamefont {M.}~\bibnamefont
  {Berz}},\ }\href@noop {} {\bibfield  {journal} {\bibinfo  {journal}
  {Part.Accel.}\ }\textbf {\bibinfo {volume} {24}},\ \bibinfo {pages} {109}
  (\bibinfo {year} {1989})}\BibitemShut {NoStop}%
\bibitem [{\citenamefont {Enge}(1964)}]{Enge_fringe}%
  \BibitemOpen
  \bibfield  {author} {\bibinfo {author} {\bibfnamefont {H.~A.}\ \bibnamefont
  {Enge}},\ }\href@noop {} {\bibfield  {journal} {\bibinfo  {journal}
  {Rev.Sci.Instrum.}\ }\textbf {\bibinfo {volume} {35}},\ \bibinfo {pages}
  {278} (\bibinfo {year} {1964})}\BibitemShut {NoStop}%
\bibitem [{\citenamefont {Forest}\ and\ \citenamefont
  {Milutinovic}(1988)}]{Forest_fringe}%
  \BibitemOpen
  \bibfield  {author} {\bibinfo {author} {\bibfnamefont {E.}~\bibnamefont
  {Forest}}\ and\ \bibinfo {author} {\bibfnamefont {J.}~\bibnamefont
  {Milutinovic}},\ }\href@noop {} {\bibfield  {journal} {\bibinfo  {journal}
  {Nucl.Instrum.Meth.}\ }\textbf {\bibinfo {volume} {A269}},\ \bibinfo {pages}
  {474} (\bibinfo {year} {1988})}\BibitemShut {NoStop}%
\bibitem [{\citenamefont {Zimmermann}(2000)}]{Zimmermann_fringe}%
  \BibitemOpen
  \bibfield  {author} {\bibinfo {author} {\bibfnamefont {F.}~\bibnamefont
  {Zimmermann}},\ }\href@noop {} {\  (\bibinfo {year} {2000})},\ \bibinfo
  {note} {{CERN-SL-2000-012-AP, CERN-NUFACT-NOTE-22}}\BibitemShut {NoStop}%
\bibitem [{\citenamefont {Berz}\ and\ \citenamefont
  {Makino}(2006)}]{Berz_cosy}%
  \BibitemOpen
  \bibfield  {author} {\bibinfo {author} {\bibfnamefont {M.}~\bibnamefont
  {Berz}}\ and\ \bibinfo {author} {\bibfnamefont {K.}~\bibnamefont {Makino}},\
  }\href@noop {} {\  (\bibinfo {year} {2006})},\ \bibinfo {note}
  {{MSUHEP-060804}}\BibitemShut {NoStop}%
\bibitem [{\citenamefont {Yoon}\ \emph {et~al.}(2004)\citenamefont {Yoon},
  \citenamefont {Corbett}, \citenamefont {Cornacchia}, \citenamefont {Tanabe},\
  and\ \citenamefont {Terebilo}}]{speardipole_Yoon}%
  \BibitemOpen
  \bibfield  {author} {\bibinfo {author} {\bibfnamefont {M.}~\bibnamefont
  {Yoon}}, \bibinfo {author} {\bibfnamefont {J.}~\bibnamefont {Corbett}},
  \bibinfo {author} {\bibfnamefont {M.}~\bibnamefont {Cornacchia}}, \bibinfo
  {author} {\bibfnamefont {J.}~\bibnamefont {Tanabe}}, \ and\ \bibinfo {author}
  {\bibfnamefont {A.}~\bibnamefont {Terebilo}},\ }\href@noop {} {\bibfield
  {journal} {\bibinfo  {journal} {Nucl.Instrum.Meth.}\ }\textbf {\bibinfo
  {volume} {A523}},\ \bibinfo {pages} {9} (\bibinfo {year} {2004})}\BibitemShut
  {NoStop}%
\bibitem [{\citenamefont {Huang}\ \emph {et~al.}(2010)\citenamefont {Huang},
  \citenamefont {Safranek},\ and\ \citenamefont {Dell'Orco}}]{Huang_IPAC10}%
  \BibitemOpen
  \bibfield  {author} {\bibinfo {author} {\bibfnamefont {X.}~\bibnamefont
  {Huang}}, \bibinfo {author} {\bibfnamefont {J.}~\bibnamefont {Safranek}}, \
  and\ \bibinfo {author} {\bibfnamefont {D.}~\bibnamefont {Dell'Orco}},\
  }\href@noop {} {\bibfield  {journal} {\bibinfo  {journal} {Proceedings of
  IPAC 2010}\ ,\ \bibinfo {pages} {4626}} (\bibinfo {year} {2010})}\BibitemShut
  {NoStop}%
\bibitem [{\citenamefont {Elleaume}(1992)}]{Elleaume_kickmap}%
  \BibitemOpen
  \bibfield  {author} {\bibinfo {author} {\bibfnamefont {P.}~\bibnamefont
  {Elleaume}},\ }\href@noop {} {\bibfield  {journal} {\bibinfo  {journal}
  {EPAC'92}\ ,\ \bibinfo {pages} {661}} (\bibinfo {year} {1992})}\BibitemShut
  {NoStop}%
\bibitem [{\citenamefont {Wu}\ \emph {et~al.}(2003)\citenamefont {Wu},
  \citenamefont {Forest},\ and\ \citenamefont {Robin}}]{Wu_id}%
  \BibitemOpen
  \bibfield  {author} {\bibinfo {author} {\bibfnamefont {Y.}~\bibnamefont
  {Wu}}, \bibinfo {author} {\bibfnamefont {E.}~\bibnamefont {Forest}}, \ and\
  \bibinfo {author} {\bibfnamefont {D.}~\bibnamefont {Robin}},\ }\href@noop {}
  {\bibfield  {journal} {\bibinfo  {journal} {Phys.Rev.}\ }\textbf {\bibinfo
  {volume} {E68}},\ \bibinfo {pages} {046502} (\bibinfo {year}
  {2003})}\BibitemShut {NoStop}%
\bibitem [{\citenamefont {Bahrdt}\ and\ \citenamefont
  {Wustefeld}(2011)}]{Bahrdt_GF}%
  \BibitemOpen
  \bibfield  {author} {\bibinfo {author} {\bibfnamefont {J.}~\bibnamefont
  {Bahrdt}}\ and\ \bibinfo {author} {\bibfnamefont {G.}~\bibnamefont
  {Wustefeld}},\ }\href@noop {} {\bibfield  {journal} {\bibinfo  {journal}
  {Phys.Rev.ST Accel.Beams}\ }\textbf {\bibinfo {volume} {14}},\ \bibinfo
  {pages} {040703} (\bibinfo {year} {2011})}\BibitemShut {NoStop}%
\bibitem [{\citenamefont {{Brookhaven National Labortary}}(2013)}]{nsls2_NEXT}%
  \BibitemOpen
  \bibfield  {author} {\bibinfo {author} {\bibnamefont {{Brookhaven National
  Labortary}}},\ }\href {http://www.bnl.gov/nsls2/project/NEXT/} {\enquote
  {\bibinfo {title} {{NSLS-II NEXT Project}},}\ } (\bibinfo {year}
  {2013})\BibitemShut {NoStop}%
\bibitem [{\citenamefont {Chubar}\ \emph {et~al.}(1998)\citenamefont {Chubar},
  \citenamefont {Elleaume},\ and\ \citenamefont {Chavanne}}]{Chubar_radia}%
  \BibitemOpen
  \bibfield  {author} {\bibinfo {author} {\bibfnamefont {O.}~\bibnamefont
  {Chubar}}, \bibinfo {author} {\bibfnamefont {P.}~\bibnamefont {Elleaume}}, \
  and\ \bibinfo {author} {\bibfnamefont {J.}~\bibnamefont {Chavanne}},\
  }\href@noop {} {\bibfield  {journal} {\bibinfo  {journal} {J.Synchrotron
  Radiat.}\ }\textbf {\bibinfo {volume} {5}},\ \bibinfo {pages} {481} (\bibinfo
  {year} {1998})}\BibitemShut {NoStop}%
\bibitem [{\citenamefont {Leemann}\ \emph {et~al.}(2009)\citenamefont
  {Leemann}, \citenamefont {Andersson}, \citenamefont {Eriksson}, \citenamefont
  {Lindgren}, \citenamefont {Wallen} \emph {et~al.}}]{Leemann_max4}%
  \BibitemOpen
  \bibfield  {author} {\bibinfo {author} {\bibfnamefont {S.}~\bibnamefont
  {Leemann}}, \bibinfo {author} {\bibfnamefont {A.}~\bibnamefont {Andersson}},
  \bibinfo {author} {\bibfnamefont {M.}~\bibnamefont {Eriksson}}, \bibinfo
  {author} {\bibfnamefont {L.-J.}\ \bibnamefont {Lindgren}}, \bibinfo {author}
  {\bibfnamefont {E.}~\bibnamefont {Wallen}},  \emph {et~al.},\ }\href@noop {}
  {\bibfield  {journal} {\bibinfo  {journal} {Phys.Rev.ST Accel.Beams}\
  }\textbf {\bibinfo {volume} {12}},\ \bibinfo {pages} {120701} (\bibinfo
  {year} {2009})}\BibitemShut {NoStop}%
\end{thebibliography}%

\end{document}